\documentclass[sigconf]{acmart}
\renewcommand\footnotetextcopyrightpermission[1]{} 
\acmYear{2021}\copyrightyear{2021}
\setcopyright{rightsretained}
\acmConference[WiSec '21]{14th ACM Conference on Security and Privacy in Wireless and Mobile Networks}{June 28--July 2, 2021}{Abu Dhabi, United Arab Emirates}
\acmBooktitle{14th ACM Conference on Security and Privacy in Wireless and Mobile Networks (WiSec '21), June 28--July 2, 2021, Abu Dhabi, United Arab Emirates}
\acmPrice{15.00}
\acmDOI{10.1145/3448300.3468256}
\acmISBN{978-1-4503-8349-3/21/06}

\usepackage{glossaries}
\usepackage{siunitx}
\usepackage{caption}
\usepackage{subcaption}
\usepackage[capitalise]{cleveref}

\glsdisablehyper

\newacronym{dhcp}{DHCP}{Dynamic Host Configuration Protocol}
\newacronym{ntp}{NTP}{Network Time Protocol}
\newacronym{tcp}{TCP}{Transmission Control Protocol}
\newacronym{udp}{UDP}{User Datagram Protocol}
\newacronym{crc}{CRC}{Cyclic Redundancy Check}

\newacronym{rss}{RSS}{Received Signal Strength}
\newacronym{mimo}{MIMO}{Multiple Input Multiple Output}
\newacronym{ofdm}{OFDM}{Orthogonal Frequency Division Multiplexing}
\newacronym{dsss}{DSSS}{Direct Sequence Spread Spectrum}
\newacronym{css}{CSS}{Chirp Spread Spectrum}
\newacronym{fsk}{FSK}{Frequency Shift Keying}
\newacronym{snr}{SNR}{Signal-to-Noise Ratio}
\newacronym{sinr}{SINR}{Signal to interference plus noise ratio}
\newacronym{aoa}{AoA}{Angle of Arrival}
\newacronym{ssc}{SSC}{Secret Spreading Code}
\newacronym{ci}{CI}{Constructive Interference}
\newacronym{cfo}{CFO}{Carrier Frequency Offset}
\newacronym{cpo}{CPO}{Carrier Phase Offset}
\newacronym{sf}{SF}{Spreading Factor}

\newacronym{rra}{RRA}{Record-and-Replay-Attack}
\newacronym{dd}{DD}{Distance-Decreasing}
\newacronym{aua}{AuA}{Area under Attack}

\newacronym{gnss}{GNSS}{Global Navigation Satellite System}
\newacronym{gps}{GPS}{Global Positioning System}
\newacronym{nmea}{NMEA}{National Marine Electronics Association}
\newacronym{nasa}{NASA}{National Aeronautics and Space Administration}
\newacronym{agps}{A-GPS}{Assisted/Augmented GPS}
\newacronym{scer}{SCER}{Security Code Estimation and Replay}
\newacronym{nma}{OS-NMA}{Open Service Navigation Message Authentication}
\newacronym{sis}{SIS}{Signal In Space}
\newacronym{raim}{RAIM}{Receiver autonomous integrity monitoring}
\newacronym{pvt}{PVT}{Position-Velocity-Time}

\newacronym{sdr}{SDR}{Software Defined Radio}
\newacronym{sdrs}{SDR}{Software Defined Radios}
\newacronym{ots}{OTS}{Off-the-Shelf}
\newacronym{rt}{RT}{Real-Time}
\newacronym{ide}{IDE}{Integrated Development Environment}
\newacronym{fifo}{FIFO}{First-In-First-Out}
\newacronym{sma}{SMA}{SubMiniature Version A}
\newacronym{rf}{RF}{Radio Frequency}

\newacronym{foi}{FoI}{Feature of Interest}

\begin{document}

\title{DEMO: Relay/Replay Attacks on GNSS signals}

\author{Malte Lenhart}
\orcid{0000-0002-8846-8657}
\affiliation{
  \institution{Networked Systems Security Group \\ KTH Royal Institute of Technology}
  \streetaddress{} 
  \city{Stockholm}
  \country{Sweden}
}
\email{lenhart@kth.se}

\author{Marco Spanghero}
\orcid{0000-0001-8919-0098}
\affiliation{
	\institution{Networked Systems Security Group \\ KTH Royal Institute of Technology}
	\streetaddress{Isafjordsgatan 26, 16440 Kista}
	\city{Stockholm}
	\country{Sweden}
}
\email{marcosp@kth.se}

\author{Panagiotis Papadimitratos}
\orcid{0000-0002-3267-5374}
\affiliation{
	\institution{Networked Systems Security Group \\ KTH Royal Institute of Technology}
	\streetaddress{Isafjordsgatan 26, 16440 Kista}
	\city{Stockholm}
	\country{Sweden}
}
\email{papadim@kth.se}

\begin{abstract}
	\glspl{gnss} are ubiquitously relied upon for positioning and timing.  
	Detection and prevention of attacks against \gls{gnss} have been researched over the last decades, but many of these attacks and countermeasures were evaluated based on simulation.
	This work contributes to the experimental investigation of GNSS vulnerabilities, implementing a relay/replay attack with off-the-shelf hardware. Operating at the signal level, this attack type is not hindered by cryptographically protected transmissions, such as Galileo's \gls{nma}.
	The attack we investigate involves two colluding adversaries, relaying signals over large distances, to effectively spoof a GNSS receiver. We demonstrate the attack using off-the-shelf hardware, we investigate the requirements for such successful colluding attacks, and how they can be enhanced, e.g., allowing for finer adversarial control over the victim receiver. 
\end{abstract}



\maketitle
\pagestyle{plain}

\section{Introduction}
Several critical applications rely on \gls{gnss} for their functionality, from autonomous navigation to sensitive infrastructure. However, civilian \gls{gnss} is inherently vulnerable to attacks against its availability (i.e., jamming) or its integrity (i.e., spoofing, forging \gls{gnss} signals to induce a false position or time).
Recently, cost-effective tools to mount spoofing attacks increase the risk \cite{ebinumaOsqzssGpssdrsim2020}. Defensive techniques in \gls{gnss} receivers are rare \cite{psiakiGNSSSpoofingDetection2016}  and often limited to jamming awareness; thus, potentially leaving millions of devices exposed to adversarial manipulation.

Jamming is directly observable and well documented \cite{g.buesnelThousandsGNSSJamming2020}; spoofing incidents are complex to detect and attribute. Russia reportedly deployed spoofing to prevent drones flying near high ranking officials in
safety-critical events \cite{UsOnlyStars}. In the research literature, spoofing was used against drones to circumvent geo-fencing and enforce landings \cite{kernsUnmannedAircraftCapture2014}. Spoofing a yacht navigation system caused the crew to adjust the course as intended by the researchers \cite{psiakiProtectingGPSSpoofers2016}.

Open-sourcing powerful attack methods is problematic (e.g., misuse of research and development tools lowers the bar for potential adversaries), but such tools are important in evaluating new countermeasures for \gls{gnss} receivers. Various detection schemes have been proposed \cite{psiakiGNSSSpoofingDetection2016, schmidtSurveyAnalysisGNSS2016}. However, without access to spoofing devices, most works are evaluated by simulation or against a set of well-defined spoofing recordings \cite{humphreysTexasSpoofingTest2012}. Commercial spoofers - if available - constitute a significant investment \cite{OroliaComOnline}, leaving only high-budget, or knowledgeable and dedicated groups in the position to acquire or build advanced spoofing systems.

Attacks can be carried out through (the misuse of) simulation (spoofing, i.e., creation of \gls{gnss} signals from publicly available data) or \textit{meaconing} (replaying recorded signals). 
Cryptographic protection can thwart spoofed/simulated GNSS transmissions but it cannot alone defend against replay attacks. 
The attacker can force the victim receiver to lock on to the adversarial signals (and thus control the victim's position/time): it can either cause a loss of lock on the legitimate signals (e.g., by jamming the victim), or by synchronously lifting-off the victim's lock from the legitimate signal. More specifically, the attacker synchronizes forged signals to the legitimate ones and gradually increases power; once both signals match, the victim continues tracking the higher-power adversarial signals \cite{humphreysAssessingSpoofingThreat2008}.

A \gls{gps} simulator was released as open source in 2015 \cite{ebinumaOsqzssGpssdrsim2020}, providing researchers a tool to test countermeasures. It is limited to \gls{gps} signals and cannot act as a synchronized spoofer. We do not expect to encounter it in realistic attacks; even though it has been adopted for self-spoofing \cite{CheatingPokemonGo2016}. 
A portable "receiver-spoofer" capable of synchronous lifting-off is presented in \cite{humphreysAssessingSpoofingThreat2008}. Signals are generated by simulation and synchronized by observation of legitimate signals. 
A record-and-replay attack investigation is presented in \cite{blumInvestigationVulnerabilityMobile2019}. 
In principle simpler than spoofing, such an attack is likely to be mounted also by less knowledgeable adversaries, thus increasing its impact.
It is however limited in terms of portability and lacks real-time replaying capabilities.

This motivates this work: we investigate the risk posed by real-time capable, long-distance replaying attackers, based on off-the-shelf hardware. We establish the attack feasibility, associated threat, and provide a framework for evaluating countermeasures.

\section{Networked Replaying Spoofer}
One advantage of meaconing is the ability to operate on encrypted and authenticated signals. Such attacks are relatively simple and potentially agnostic to the future evolution of \gls{gnss} signals. Furthermore, by changing the meaconer center frequency we can selectively target a subset of \gls{gnss} bands and constellations.

\begin{figure}
	\centering
    \includegraphics[width=0.47\textwidth]{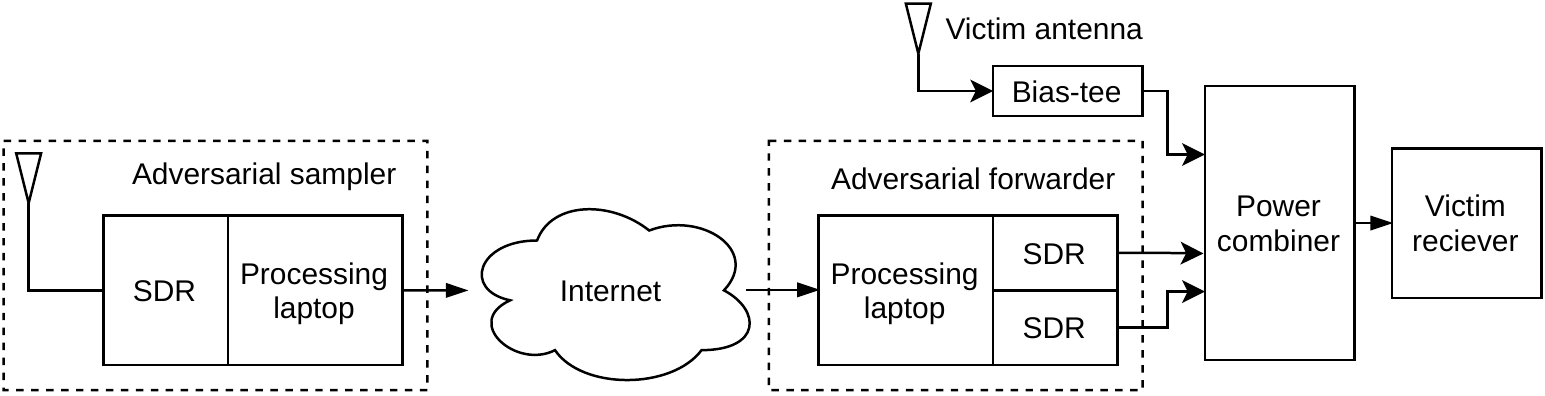}
    \caption{Block diagram of the colluding adversaries. Both adversarial nodes have a \gls{sdr}, the forwarder is connected to the victim \gls{gnss} receiver via a power combiner. To force a loss of lock at the victim receiver, the adversary jams the appropriate frequencies.}
    \label{fig:networked-hw}
	\Description{System block diagram depicting adversarial sampler, adversarial forwarder and the victim receiver. The adversarial sampler has a processing laptop connected to a software defined radio with an actively fed global satellite navigation systems antenna to record legitimate signals. The adversarial sampler sends the recorded signals over the internet to the adversarial forwarder, also equipped with a processing laptop and a software defined radio. The adversarial forwarder replays the signal, which is then combined with the legitimate signal from a second active antenna before being fed to the victim receiver.}
\end{figure}

The networked relay/replay attacker, illustrated in \cref{fig:networked-hw} consists of two colluding adversaries, with one entity (the adversarial sampler) recording legitimate \gls{gnss} signals, and the other (the adversarial forwarder) replaying them at a different location. 
In order for the attack to be successful (i.e., having the victim lock upon the adversarial signals and derives false position and time, due to the attack), the network bandwidth and latency have to be sufficient. 

\subsection{Network Prerequisites}
\label{sec:network}
The network bandwidth required to replay the recorded signal depends on sample rate and quantization bits used by the adversarial sampler ($ dataRate = sampleRate * quantizationBits * 2$), taking into account the in-phase and quadrature components. 
Our preliminary investigation shows that we can achieve the lowest data rate of \SI{31,88}{Mb/s} at a sample rate of \SI{1}{MHz} and 16-bit quantization. 
The adversarial nodes are connected over TCP sockets, as UDP proved too lossy in test transmissions, not allowing a stable replay. Therefore, the choice of sampling parameters needs to be conservative with respect to the network bandwidth.

Sufficient bandwidth to act as an adversarial forwarder, highly dependent on the  consumer-grade Internet in the area, is often available \cite{WorldwideBroadbandSpeed}. Upload, typically lower than download, is the limiting factor for the adversarial sampler. Such limitations will be increasingly less significant with the roll-out of fast 5G cellular network.

\subsection{Experiment Setup}
\label{sec:setup} 
\begin{figure}
    \centering
    \hfill
    \begin{subfigure}[b]{0.162\textwidth}
        \centering
        \includegraphics[width=0.8\linewidth]{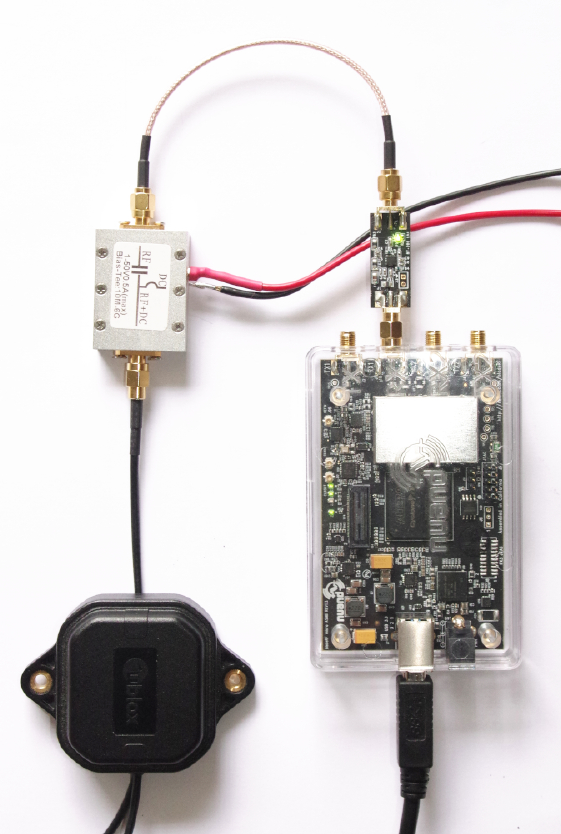}
        \caption{Adv. sampler}
    \end{subfigure}
    \hfill
    \begin{subfigure}[b]{0.27\textwidth}
        \centering
        \includegraphics[width=0.65\linewidth]{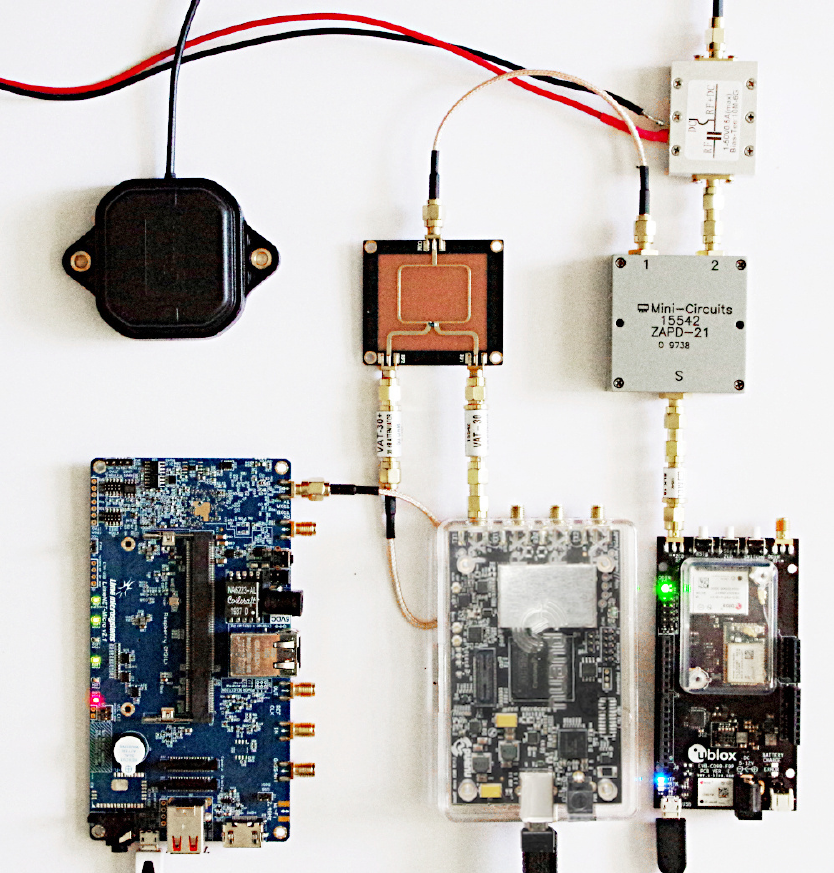}
        \caption{Adv. forwarder}
    \end{subfigure}
    \hfill
    \caption{(a) The adversarial sampler \gls{sdr} is connected to a Low-Noise-Amplifier, a bias-tee and an active \gls{gnss} antenna. (b) The victim receives the combination of legitimate \gls{gnss} signal, adversarial jamming and replayed signal.}
    \label{fig:hw-photo}
    \Description{Photo of the hardware as described in the figure caption. }
\end{figure}

\cref{fig:hw-photo} shows the two colluding adversarial nodes, equipped with \textit{bladeRF 2.0} \glspl{sdr} and connected over the Internet. The adversarial forwarder has a second \gls{sdr} to jam the \gls{gps} bands tracked by the victim. 
The \textit{u-Blox F9P} victim receiver is connected to the the adversarial forwarder and an active antenna to track legitimate signals. 
The adversarial sampler receives the legitimate \gls{gnss} signal, compresses it, and relays it to the adversarial forwarder. The forwarder re-transmits it to the victim, that logs the perceived \gls{pvt} solution. 

To cause our Multi-\gls{gnss} receiver victim to loose lock to the legitimate \gls{gnss} signals, the adversary jams \gls{gps} L1 and L2 bands (by transmitting Gaussian noise) for a time period in which it can be assumed that the victim looses track of all satellites.  
All bands have to be jammed, so that the victim cannot keep tracking satellites in an undisturbed band, leading to rejection of the replayed signal. At this stage, the replay is limited to one band, therefore the other bands are consecutively jammed after meaconing is initiated. 

\subsection{Evaluation}
\label{sec:eval}
To evaluate relaying/replaying over a long distance, the adversarial sampler is located in Stockholm at the KTH Campus in Kista, the adversarial forwarder is positioned in Germany at a distance of \SI{1100}{km}\footnote{The reverse direction was also tested to assess suitability of consumer grade Internet for this attack. In the first attempt, the upload speed (\SI{11}{Mb/s}) did not suffice, leading to congestion and a non-continuous signal at the victim. A repetition in a place with an upload of \SI{49}{Mb/s} allowed misleading the Sweden-based victim that its location was in Germany. This shows that a change in location and a Internet provider can determine the success of the attack.}.
We observed that short-term network congestion leads to a brief loss of satellite tracking at the victim, followed by resuming tracking. Such gaps can be compared to natural outages in areas with few or no visible satellites. If the adversarial sampler - adversarial forwarder bandwidth is insufficient for longer periods, the attack fails.

In the first experiment, we attacked the victim receiver that has no prior lock to legitimate signals (i.e., 'cold start'). After a brief interval, the receiver accepted the spoofed signals over the weaker legitimate ones. 
In the second experiment, the receiver is locked onto legitimate signals, before jamming is initiated. We observe the victim loosing track of all satellites and then we start replaying. 

\begin{figure*}
     \centering
     \begin{subfigure}[t]{0.25\textwidth}
         \centering
         \includegraphics[width=\linewidth]{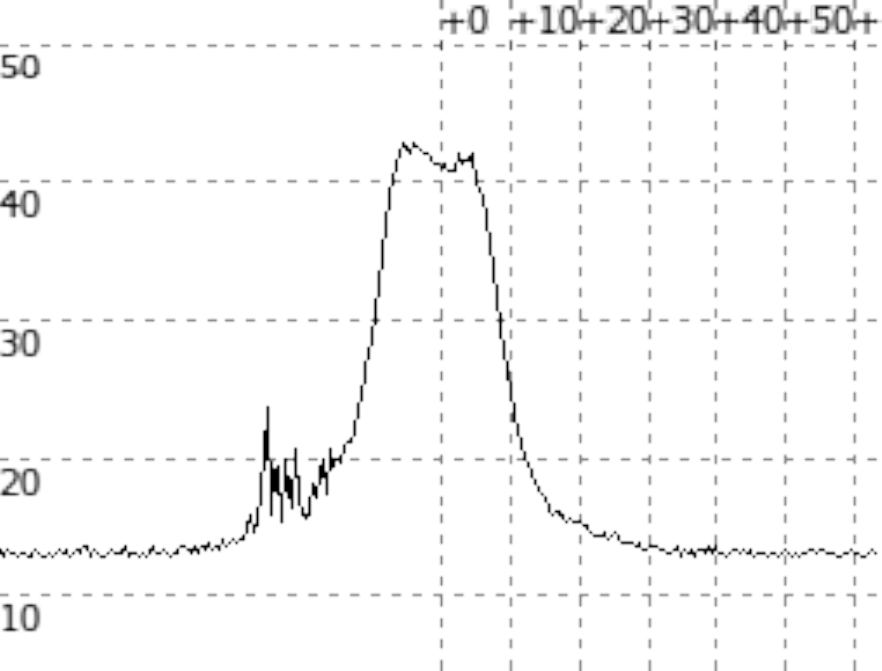}
         \caption{Legitimate signal spectrum at the victim receiver}
         \label{fig:original-spectrum}
     \end{subfigure}
     \hfill
     \begin{subfigure}[t]{0.25\textwidth}
         \centering
         \includegraphics[width=\linewidth]{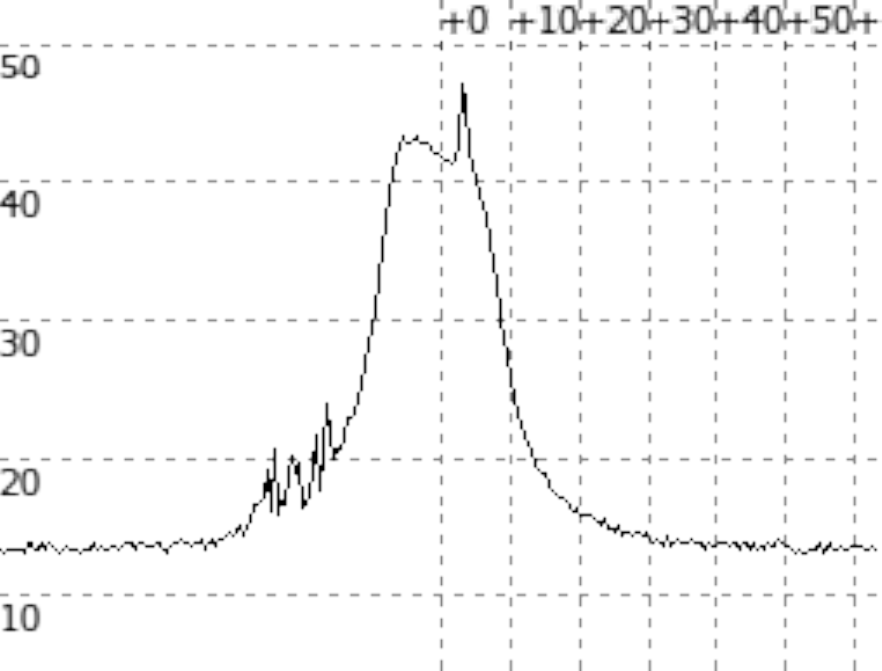}
         \caption{Spectrum of combined legitimate and replayed signal}
         \label{fig:active-ant-replay}
     \end{subfigure}
     \hfill
     \begin{subfigure}[t]{0.25\textwidth}
         \centering
         \includegraphics[width=\linewidth]{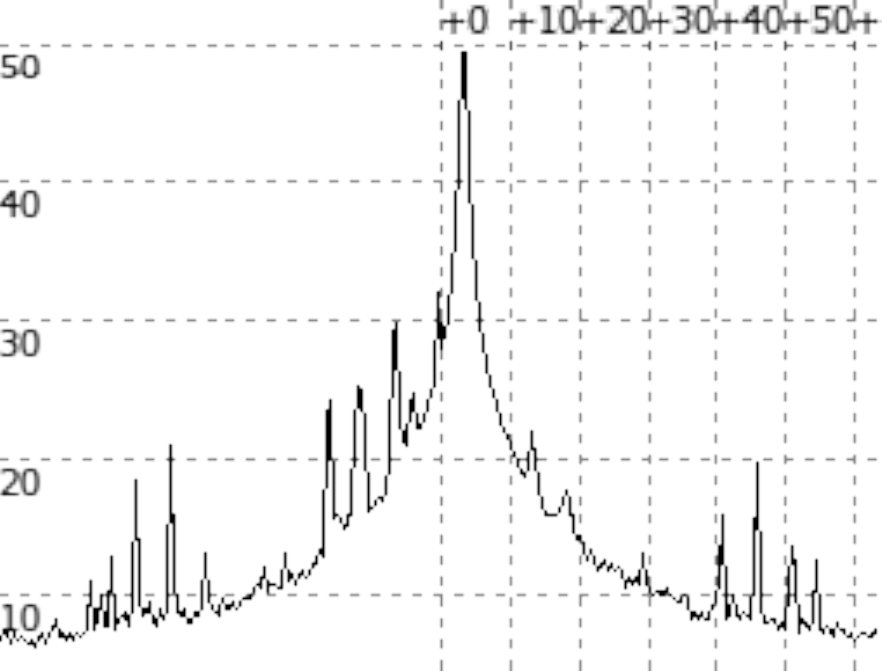}
         \caption{Effect of jamming on legitimate spectrum.}
         \label{fig:passive-ant-feed}
     \end{subfigure}
        \caption{Victim receiver spectrogram. Center frequency is $f_c = \SI{1.57542}{\giga\hertz}$, power in \si{\decibel}. (a) Legitimate signal. (b) Addition of the replayed signal causes a small spike (top right) in the spectrum, due to higher power and narrow bandwidth of the replayed signal. (c) Jamming changes the perceived spectrum drastically, usable for attack detection. }
        \label{fig:signals}
        \Description{Three spectrogram views. Without adversarial influence, a broad plateau is visible around the center frequency. Addition of adversarial meaconing signal shows in a small spike at the outer edges of the plateau. Adversarial narrow-band jamming manifests in multiple 5 to \SI{8}{dB} spikes, as well as a changed perception in the sidebands.}
\end{figure*}

When switching from jamming to replaying in the L1 band, it takes a few minutes until the receiver accepts the replayed signals and is mislead to be at the adversarial location\footnote{The receiver locks back onto legitimate signals a few minutes after the attack ends.}. We observed that in some instances the victim perceives the spoofed satellites but does not lock onto them. In this cases a brief second jamming leads to an immediate lock on in the victim once replaying resumes. 
Legitimate and relayed/replayed signal spectrograms are depicted in \cref{fig:signals}. The addition of replayed signals slightly distorts the power density at center frequency due the bandwidth limitation of the replayed signal. This also explains why the satellites on \gls{gps} L2 frequency are missing in the receiver side leading to a larger position error. Jamming changes the  perceived spectrum drastically, which among others can be monitored by receivers to detect jamming. 

\section{Conclusion}
\label{sec:conclusion}
This demonstration shows that long-distance replay attacks on \gls{gnss} are possible in real-time and with off-the-shelf hardware. As Galileo \gls{nma} is currently in testing phase \cite{TestsGalileoOSNMA2021}, 
our tests do not include authenticated signals. This is part of future work and we anticipate the relaying/replaying attack effect to be the same. Cryptographic methods do not prevent relaying/replaying attacks, but their combination with other detection schemes, such as drift monitoring or signal distortion \cite{psiakiGNSSSpoofingDetection2016}, can shield \gls{gnss} receivers.

Low-rate Internet and expensive hardware made this attack type infeasible in the past, but now it would be possible to setup a network of replayed location sources, streaming legitimate \gls{gnss} signals. Such a distributed setup, with several adversarial forwarders and samplers, strengthens the adversarial options.
Our experiments involved static adversarial sampler and transmitter; in future work, we will work with mobile nodes, investigating the feasibility of such attack using cellular networks.

\textbf{Ongoing work:} We are working on relaying signals at the message level. For this, we demodulate and analyze the signal at the adversarial sampler and rebuild it, based on signal parameters, at the forwarder. This significantly reduces the required network bandwidth and enables more advanced attack types, such as delaying selected satellite signals. This influences the victim's derived position directly, beyond what is possible with relaying \cite{papadimitratosProtectionFundamentalVulnerability2008}.

Replayed signals are hard to be code-phase synchronized with the legitimate signals, due to the processing and network delay. To a certain extend, this can be compensated by sophisticated distance-decreasing attacks \cite{zhangEffectsDistancedecreasingAttacks2019}, with their investigation being part of ongoing work.

\section{Demonstration Setup}
In the scope of the virtual conference, we showcase our networked relay/replay spoofing attack in a pre-recorded video.
We introduce the utilized software and hardware, and show the system state shown prior to the attack, with receivers at both ends showing the legitimate derived positions.
To start the attack, the adversarial nodes establish the unidirectional sample stream. 
The adversarial forwarder initiates jamming, and then replays the streamed signals to the victim. After a while, the victim accepts the spoofed position, thus concluding our demo.

\begin{acks}
    This work was supported by the Swedish Foundation for Strategic Research (SSF) SURPRISE project and the KAW Academy Fellow Trustworthy IoT project.
\end{acks}

\bibliographystyle{ACM-Reference-Format}

\bibliography{gnss-update}

\end{document}